\documentclass[twocolumn,nofootinbib,aps,prl,showpacs]{revtex4-1}
\usepackage{dsfont}
\usepackage{graphicx}
\usepackage{float}
\usepackage{pdfpages}
\usepackage{amsmath}
\usepackage{amsfonts}
\usepackage{hyperref}
\usepackage{fullpage}
\newcommand{\mathsym}[1]{{}}

\begin{document}
\title{What a single snapshot reveals about the future and the past of turbulent flow}
\author{Gregory Falkovich and Anna Frishman}
\affiliation{Physics of Complex Systems, Weizmann Institute of Science, Rehovot 76100 Israel }
\date{\today}
\begin{abstract}
We develop an analytic formalism and derive new exact relations that express the short-time dispersion of fluid particles via the single-time velocity correlation functions in homogeneous isotropic and incompressible turbulence. The formalism establishes a bridge between single-time Eulerian and long-time Lagrangian pictures of turbulent flows. In particular, we derive an exact formula for a short-term analog of the long-time Richardson law,  and we identify a conservation law of turbulent dispersion which is true even in non-stationary turbulence.

\end{abstract}


 \maketitle
{\bf Introduction}. There are two alternative types of description in physics: one either uses a coordinate system fixed in space 
or follows particles. In fluid mechanics, the former is called Eulerian and the latter Lagrangian \cite{Fal}. Respectively, our knowledge of turbulence roughly can be divided into two parts. On the Eulerian side, we have experimental and numerical data on the single-time (snapshot) velocity statistics: the moments of the velocity difference $\vec{u}$ measured at the distance $R$ are $\langle u^n\rangle \propto R^{\xi_n}$. As for theory, the only analytic result is $\langle u_l^3\rangle =-12\epsilon R/d(d+2)$ where $\epsilon$ is the energy spectral flux, negative in 2d and positive in 3d, $u_l=(\vec{u}\cdot \vec{R})/R$ and $d$ is space dimensionality \citep{Kolmogorov1941,Frisch}.
For $d=3$, it is Kolmogorov's $4/5$-law.  On the Lagrangian side,  the long-time growth of the inter-particle distance was inferred by Richardson from atmospheric data: $\langle R^2(t)\rangle\simeq |\epsilon| t^3$, which was neither properly observed in a controlled laboratory experiment nor derived analytically. Here and below, all averages are over an ensemble of pairs whose separation initially was the same: $\vec{R}(0)=\vec{R}_0$. Kolmogorov and Richardson laws, respectively Eulerian and Lagrangian, are apparently related by $\epsilon$ and yet it is not known if the latter is a consequence of the former.

Here we develop a formalism, which is a bridge between Eulerian and Lagrangian approaches. We derive a relation, which is a Lagrangian consequence of the Kolmogorov law and a short-time counterpart to the Richardson law. We express the Lagrangian time derivatives of the correlation functions of $\vec{u}$ and $\vec{R}$ at $t=0$ in terms of the Eulerian moments of $\vec{u}$. For short times, these determine the temporal evolution of the correlation functions by a Taylor expansion in $t$. Previously, analytic results on Lagrangian evolution were derived either for spatially trivial smooth flows at long-time limit or for temporally trivial short-correlated  Kraichnan model. We derive some new results for these two cases, as well as for the general case of spatially non-smooth temporally correlated flow which corresponds to the inertial interval of turbulence.

Of most interest is to find quantities conserved in time or at least having time derivatives all zero at $t=0$; as we shall see below, the latter does not guarantee the former. The only fact known before about two particles is that $\left\langle [R(t)]^{-d} \right\rangle$ does not change at $t\to\infty$ in a spatially smooth flow (in the viscous but not inertial interval)
\cite{zel1984kinematic}. One may think that the time dependence saturates only in the long-time limit when starting conditions are forgotten. However, here we show
that $\left\langle R^{-d} \right\rangle$ is an all-time genuine integral of motion for statistically isotropic spatially smooth random flow, even with a non-stationary statistics. To turn time derivatives into zero we shall use the simple mathematical fact:
\begin{equation}\nabla_iR_iR^{-d}=\delta(R)\ .\label{simple}\end{equation}
Using that it is zero for $R\not=0$, we find the general conserved quantity: $\bigl\langle u^{a-m}R^{-d-a-m}(\vec{u} \cdot \vec{R})^m \bigr\rangle$.

For the Kraichnan model with spatially non-smooth velocities we show that the moment, which generalizes  $\left\langle R^{-d} \right\rangle$, has all time derivatives zero at $t=0$, although it is not constant in time. 
For real turbulence in the inertial interval (spatially non-smooth yet temporally finite-correlated), we find that a direct analog is $\left\langle R^{2-\xi_2-d} \right\rangle$. It has the first two derivatives equal to zero, but not the third one, showing what little traces of the conservation are left away from the limiting cases of velocities either spatially smooth or temporally white. This moment is nonetheless of much physical interest as both a short-time analog of the Richardson law, whose evolution is determined by the energy flux, and as a measure of the irreversibility of particle dispersion.

{\bf Statistics of the distances}. Let us first calculate the time derivatives of correlation functions involving only $R$ using the pdf
\begin{equation}{\cal P}(R,t;R_0,0)=\left\langle \delta \left(\vec{R}-\vec{R}_0-\int_0^t\!
 \vec{u}(t')dt'
\right) \right\rangle\ .\label{P}
\end{equation}
The integral is over Lagrangian trajectories and averaging here and everywhere is over $\vec{u}$.
To calculate time derivatives at $t=0$ of any mean, 
\begin{equation}\left\langle F\bigl(R(t)\bigr)\right\rangle= \!\int\! F(R){\cal P}(R,t;R_0,0)dR\ ,\label{F}
\end{equation}
we first differentiate  the $\delta$-function (and velocities), then set  $t=0$ turning the $\delta$-function into $\delta(R-R_0)$:
\begin{eqnarray}\!\!\!\!\left. \frac{\partial^n\! \langle F (R)  \rangle }{\partial t^n}\! \right|_{t=0}\!\! =\! \left\langle \frac{d^n F (R)}{dt^n} \right\rangle_{t=0} \!\!\! = \left\langle \frac{d^n}{dt^n} \right\rangle\! F (R_0)\,,\label{general0} \end{eqnarray}
where $ {d}/{dt} \equiv \partial_t+\vec{u} \cdot \nabla_R$ is implied. Here and below, unless stated otherwise, averages without an explicit specification of the time are taken at $t=0$.

When the velocity statistics is stationary,
\begin{eqnarray} && \left\langle\!  {d^2\over dt^2}\! \right\rangle  = \langle u_i\nabla_i u_j\rangle\nabla_j\,,\  \  \left\langle\!  {d^3\over dt^3}\! \right\rangle\! =\!\langle u_i\nabla_i u_j\nabla_ju_k\rangle \nabla_k\,,\nonumber\\&&\left\langle\! {d^4\over dt^4}\!\right\rangle\! = \! \langle  u_i\nabla_i u_j\nabla_ju_k\nabla_ku_l \rangle\nabla_l-\nabla_i\left\langle\!  {\partial u_i\over\partial t}{\partial u_j\over\partial t}\! \right\rangle\nabla_j \nonumber\,.\end{eqnarray}
Velocity time derivatives (acceleration etc) appear only starting from $n=4$. Therefore, in a general case the velocity snapshot contains enough information to describe evolution up to $t^4$ terms. For translation  invariant incompressible flow, the first two derivatives can be simply obtained from the velocity single-time pdf $f(\vec{u})$ assuming that the velocity difference does not change with time:
\begin{eqnarray} {\cal P}(\vec{R},t;\vec{R}_0,0)
=|t|^{-d}f [({{\vec{R}-\vec{R}_0})/{t}}]\ .\nonumber
\end{eqnarray}
This ballistic regime was first considered in \cite{Bat50}, and it was recently suggested to be relevant for many physical situations~\cite{bourgoin2006role}.

Consider now general scalar correlation functions of the form $ \bigl\langle F \bigl(\vec{u}(t),\vec{R}(t)\bigr)\bigr\rangle$ which may also include  time derivatives $ {d^n\vec{u}}/{dt^n}$. Let us label the particles by the indices $1,2$ and denote Eulerian spatial coordinates by $r$.
Taking the Lagrangian time derivative,
\begin{eqnarray}\left. \frac{d}{dt}\right|_{ \vec{R}_{0,1},\vec{R}_{0,2}}\!\!=\left.\frac{\partial}{\partial t}\right|_{\vec{r}_1,\vec{r}_2}+\left.v_1^i \frac{\partial}{\partial r_1^i}\right|_{t,\vec{r}_2}+\left.v_2^i \frac{\partial}{\partial r_2^i}\right|_{t,\vec{r}_1}\,,\nonumber
\end{eqnarray}
at $t=0$, one can set $\vec{R}=\vec{R}_0$ in the term with the partial time derivative.
It then turns into an Eulerian correlation function, where the average is done over all pairs in the flow separated by the same distance.
Using incompressibility and homogeneity we rewrite the sum of spatial derivatives and obtain
\begin{eqnarray}
 \frac{\partial }{\partial t}\!\left\langle\!   F(\vec{u},\vec{R})  \right\rangle_{t=0} & =&\frac{\partial }{\partial t} \left\langle  F(\vec{u},\vec{R_0}) \right\rangle \nonumber\\&+&
\nabla_i\langle u^i  F(\vec{u}, \vec{R}_0)\rangle.\label{flux_relation0}
\end{eqnarray}

We apply the formalism first to a spatially smooth random flow with isotropic but not necessarily stationary statistics. In this case, $\xi_n=n$ i.e. the relative velocity is proportional to the distance, $\vec{u}=\hat\sigma(t)\vec{R}(t)$, where
$\hat\sigma(t)$ is a random traceless matrix with isotropic statistics. Such is the case, for example, in the viscous interval of turbulence, where the separation between the particles is much smaller than the viscous scale, for elastic turbulence and many other cases related to dynamical chaos. It has been shown by  Zel'dovich et al. \cite{zel1984kinematic} that $\left\langle [R(t)]^{-d} \right\rangle$  does not change at $t\to\infty$ in a steady flow, using the formalism of Furstenberg \cite{furstenberg1963noncommuting}. Another proof under the same conditions is in \cite{falkovich2001particles}. Let us prove that all the time derivatives at $t=0$ are zero for $F(R)=R^{-d}$. This is a consequence of $\left\langle d^n R^{-d}/dt^n \right\rangle=0$ when taken at $R=R_0$ without additionally setting $t=0$ in $\hat\sigma(t)$.
To show this we note that the first time derivative at $R=R_0$ is zero by virtue of isotropy:
\begin{equation}
\left\langle {d R^{-d}}/{dt}\right\rangle =\left\langle \sigma_{ij}(t)\right\rangle R_0^i R_0^j R_0^{-d-2}=0
\end{equation}
We can now use induction and express the $n+1$  time derivative at $R=R_0$ similarly to (\ref{flux_relation0}),
\begin{eqnarray}
\left\langle\!\frac{d^{n+1}}{dt^{n+1}} R^{-d}\right\rangle =\partial_t \left\langle\frac{d^{n}}{dt^{n}} R^{-d}\right\rangle +\nabla_i\left\langle u^i\frac{d^{n}}{dt^{n}} R^{-d}\right\rangle\,.\nonumber
\end{eqnarray}
On the right side, the first term is zero by the induction hypothesis, and the second term is zero because of (\ref{simple}) and $\left\langle u^i {d^{n}R^{-d}}/{dt^{n}}\right\rangle\propto R_0^i R_0^{-d}$ at $R=R_0$. An alternative derivation is given below by (\ref{time_der_zero}-\ref{condition}) for the stationary case. Our simple proof shows that \begin{equation}
\label{R^-d}
\left\langle [R(t)]^{-d} \right\rangle=R_0^{-d}\,,
\end{equation} for all $t$ even for a time-dependent statistics. The universality of this statistical conservation law follows from the simple dynamical statement: in every realization of an incompressible spatially smooth flow, the integral of $[R(t)]^{-d}$ over the directions of the initial vector, $\vec{R}_0$, is constant in time. That statement can actually be found by taking a closer look at the argument (not only result) of \cite{zel1984kinematic}. For  isotropic flows this integral can be interpreted as an average over the angular degrees of freedom, which gives (\ref{R^-d}).
Understanding that (\ref{R^-d}) holds for all times, even in the case of decaying turbulence, opens the door for its experimental verification, lacking so far.


{\bf Non-smooth velocity}. Now we ask if a similar conservation law exists in a spatially non-smooth case. The only known results are for the Kraichnan model where the velocity is statistically stationary, homogenous and delta-correlated in time: $  \langle u_i(t)u_j(t')  \rangle=  D_{ij}(\vec{R})\delta(t-t')$. The pdf satisfies the equation $\partial_t {\cal P}=\nabla_iD_{ij}\nabla_j{\cal P}={\cal M}{\cal P} $, see e.g. \cite{falkovich2001particles}. For an incompressible statistically isotropic flow, $D_{ij}(\vec{R})= R^{-\gamma}[(d+1-\gamma)\delta_{ij}R^2-(2-\gamma)R_iR_j]$, where $\gamma$ is thus the measure of velocity non-smoothness, and  time scales as $R^\gamma$. The role played by $\left\langle R^{-d} \right\rangle$ in a smooth case is now assumed by $\left\langle R^{\gamma-d} \right\rangle$, which follows from  ${\cal M} R^{\gamma-d}=\delta(R)$. Indeed, this means that the first time derivative at any $t$ is proportional to the probability of two particles initially separated by a distance $R_0$ to come together at time $t$ \cite{falkovich2001particles}: \begin{equation}
\frac{\partial}{\partial t}\left\langle R^{\gamma-d} \right\rangle = {\cal P}(0,t;R_0,0),
\end{equation}
Using the pdf from  \cite{balkovsky1998instanton,falkovich2001particles}, we derive:
\begin{equation}
{\cal P}(0,t;R_0,0) \propto  {R_0^{d-1}}{|t|^{-d/\gamma}}\exp\left(- {dR_0^{\gamma}}/{\gamma^2|t|} \right).
\end{equation}So, while $\left\langle R^{\gamma-d} \right\rangle$ is not conserved in the Kraichnan model, all its time derivatives are zero at $t=0$ since ${\cal P}(0,t;R_0,0)$ has a substantial singularity.
Time derivatives in the Kraichnan model differ from (\ref{general0}), where we first set $t=0$ and only then average over velocities. The two procedures commute for finite correlated velocities but not for $\delta$-correlated ones.

Let us see if there is any special moment $F(\vec{R})=R^b$ in a real turbulence where velocities have  finite temporal correlations. From now on we consider stationary statistics. The first time derivative is zero for any $b$ due to isotropy, then, assuming in addition incompressibility and translational invariance,
\begin{eqnarray}
&&\left\langle  \frac{ d^{2}}{dt^{2}}{R^b\over R_0^b}   \right\rangle =a_2 b {\left\langle u_l^{2} \right\rangle\over R_0^2}\ , \nonumber \\&&\left\langle   \frac{ d^{3}}{dt^{3}}{R^b\over R_0^b}   \right\rangle =a_3(a_3+b)\frac{b}{2}  {\left\langle u_l^{3}\right\rangle\over R_0^3}\,.\label{third_R_derivative}
\end{eqnarray}
where $a_k=d-k+\xi_k+b$.
For all moments but one, the second derivative is non-zero and short-time evolution is quadratic in time i.e. ballistic.

The exceptional moment $R^{2-\xi_2-d}$, which has both first and second derivatives zero, is a direct analog of $R^{\gamma-d}$ in the Kraichnan model.
The similarity between these two cases is related to the fact that the operator $\langle {d^2/ dt^2}\rangle =\nabla_i\langle u_iu_j\rangle\nabla_j\equiv {\cal\tilde{M}}$ has the same structure as the operator ${\cal M}\equiv \nabla_i D_{ij}\nabla_j$, which determines the time evolution of the distance pdf in the Kraichnan model. Moreover, zero time derivatives at $t=0$ of $\left\langle R(t)^{2-\xi_2-d} \right\rangle$ also originate from ${\cal\tilde{M}} R^{2-\xi_2-d}=\delta(R)$. However, the third time derivative (\ref{third_R_derivative}) of  $R^{2-\xi_2-d}$ is non-zero, in distinction from the Kraichnan model.

Similarity to the Kraichnan model may tempt one to conjecture that $ \left\langle {d^2} R^{2-\xi_2-d}/ {dt^2}\right\rangle \propto {\cal P}(0,t;R_0,0)$ for finite correlated velocities. To show that this is not the case, note that $\partial_t^n{\cal P}(R,t;R_0,0)$ at $t=0$ is proportional to $\delta(\vec{R}-\vec{R}_0)$ that is zero for $\vec{R}\not=\vec{R}_0$. Thus, ${\cal P}(0,t;R_0,0)$ has a substantial singularity at $t=0$ like in the Kraichnan model.
Since  $ \left\langle {d^2} R^{2-\xi_2-d}/ {dt^2}\right\rangle $ has a non-zero first derivative, then it cannot be proportional to ${\cal P}(0,t;R_0,0)$ in finite-correlated flows. Indeed, it is quite general that a probability to cross a finite distance (from $R_0$ to $0$) has a substantial singularity at $t=0$. On the other hand, non-analyticity of the moment evolution $\langle R^{b}(t)\rangle$ is an artefact of the delta-correlated model and does not take place for a finite-correlated flow.

Let us now describe the short-time evolution of  $\left\langle [R(t)]^{2-\xi_2-d} \right\rangle$ using the energy flux relations in 2d and 3d combined with equation (\ref{third_R_derivative})
and the Kolmogorov scaling $\xi_3=1$ and $\xi_2=2/3$. We get respectively for 2d and 3d:
\begin{eqnarray}
&&\left\langle \left[{R_0 \over R(t)}\right]^{2/3} \right\rangle -1=\frac{2\epsilon t^3}{27 R_0^{2}}\,,\label{2dR}
\\
&&\left\langle  \left[{R_0 \over R(t)}\right]^{5/3} \right\rangle-1= \frac{14{\epsilon} t^3 }{81R_0^{2}}\ .
\label{3dR}\end{eqnarray}
In 3d, where $\epsilon>0$ and the energy flows to small scales, growth of this negative moment means that the main contribution comes from converging pairs. In 2d, where $\epsilon<0$ and the energy flows to large scales, diverging pairs dominate and the moment decays. These  relations provide the Lagrangian consequence of the $4/5$-law and the short-time counterparts to the Richardson law.

The moment $\left\langle R^{2-\xi_2-d} \right\rangle$ thus provides an alternative way of measuring ${\epsilon}$, which is valid for short times and so should be accessible experimentally. Furthermore, the initial dynamics of this special moment is irreversible in time, unlike other moments of the separation where irreversibility is hidden by the ballistic evolution. At long times, we expect $\left\langle R^{2-\xi_2-d} \right\rangle$ to decay both in 2d and 3d, since diverging pairs should dominate the statistics. Indeed, if the pdf is a self-similar function of $R^{2}/\epsilon t^3$, which seems to be the case at least in 2d, we obtain $\langle R^{4/3-d}\rangle\propto t^{2-3d/2}$.

We thus find that in the inertial interval of turbulence (spatially non-smooth velocity with finite temporal correlations), there are no moments whose first three time derivatives turn into zero; we conclude that the velocity snapshot completely determines the short-time evolution of the statistics of the distance between fluid particles.

{\bf Velocity-distance correlation functions}. For stationary statistics, $\partial_t \langle F\bigl(\vec{u},\vec{R_0}) \bigr\rangle=0$, so that (\ref{flux_relation0}) turns into a continuity equation:
\begin{equation}
\label{flux_relation}
 \frac{\partial }{\partial t}\left\langle   F(\vec{u}(t),\vec{R}(t))  \right\rangle_{t=0} =\frac{\partial}{\partial R_0^i}\langle u^i  F(\vec{u}, \vec{R}_0)\rangle
\end{equation}
The particular case is the Kolmogorov law expressed in a Lagrangian language: $\partial_t\langle u^2 \rangle=\langle du^2/{dt} \rangle=\nabla_i \langle u^i u^2\rangle = -4{\epsilon}$~ \cite{falkovich2001particles,ott2000experimental}. We can now generalize
\begin{equation}
\left\langle {d u^2 R^b}/{dt} \right\rangle= -4{\epsilon} R_0^a \left(1+{b}/{d}\right).\label{dens}
\end{equation}
The initial evolution of the moments depending on the energy $u^2$ is controlled by ${\epsilon}$.
The exception is as an effective energy density, $\langle [u(t)]^2 [R(t)]^{-d} \rangle$, whose first  derivative is zero and the short-time behavior is dominated by a time reversible $t^2$ contribution.
More generally,  $\left\langle d u_l^a R^b / dt\right\rangle=c_1R_0^{b-1} \langle u_l^{a+1}\rangle$ and $\langle d^2 u_l^a R^b / dt^2\rangle=c_2(b+a c_2+a) R_0^{b-2} \langle u_l^{a+2}\rangle/(a+1)$, where $c_k=d-k+\xi_{k+a}+b$. One can show that no $a,b$ exist which turn both derivatives into zero in a non-smooth case.
We thus cannot build statistical integrals of motion compensating growth of $u$ by that of $R$.
Consider even more general form:
\begin{equation}
\label{G_def}
\!\! G(\vec{u},\vec{R}) =u^{a-m}R^{b-m}(\vec{u} \cdot \vec{R})^m
\end{equation}
with arbitrary $a,b,m$.
From our short-time perspective, if $\left\langle G(t) \right\rangle$ doesn't have all time derivatives equal to zero at the initial time, it is surely not an integral of motion.
Using stationarity and (\ref{flux_relation}), we  express the condition for time derivatives of $\left\langle G(t) \right\rangle$ to vanish:
\begin{equation}
\begin{split}
\label{time_der_zero}
\left\langle \frac{d^{n}G}{d t^{n}} \right\rangle_{t=0} = 
\nabla_{i}\left\langle u_i \frac{d^{n-1}}{dt^{n-1}} G \right\rangle =0.
\end{split}
\end{equation}
The last equality would be satisfied by virtue of (\ref{simple}) if
$\langle u_i {d^{n-1}G}/{dt^{n-1}}\rangle ={Const }R_0^i{R_0^{-d}}$.
If $b$ is not a positive integer, that correlation function contains the term
\begin{equation}
\label{velocity_scaling}
\left\langle u^i u^{n-1+a}\right\rangle R_0^{b-n+1} \propto R_0^{b-n+\xi_{n+a}}R_0^i\
\end{equation}
for any $n$.
Assuming that all terms in that function have the same scaling, we
obtain the condition to satisfy (\ref{time_der_zero}):
\begin{equation}\label{condition}
b-n+\xi_{n+a}=-d.
\end{equation}
It is valid for all $n$ only for the smooth case $\xi_n=n$ and $b+a=-d$, which generalizes (\ref{R^-d}). We thus found a whole family of quantities whose time derivatives are all zero at $t=0$. To claim conservation, 
we need analyticity in time, which holds when (\ref{G_def}) is an analytic function of the velocity (which is an analytic function of time).  We thus have an infinite family of conservation laws in a spatially smooth random flow. The 
entire family originates from the same dynamical statement as the special case (\ref{R^-d}).

We mention briefly the vorticity cascade in 2d, where $\xi_n=n$, 
however it is straightforward to show that the logarithmic corrections to the scaling prevent the existence of the integrals in the form (\ref{G_def}). Therefore, statistical integrals of motion for a pair of particles are found, so far, only for spatially smooth random flows (and 1d compressible random flow 
where the inter-particle distance itself is the statistical integral of motion \cite{DFTT}). For  Burgers turbulence, one can also find the analog of (\ref{dens}) with $d=1/3$. We conclude that for non-smooth flows, including
the Kolmogorov scaling $\xi_n={n}/{3}$, statistical integrals of motion that exploit basic symmetries and  (\ref{simple}) do not exist. Either the integrals appear at the long-time limit or finding them requires deeper insight into the interplay between dynamics and geometry.

The work was supported by the grants of ISF, BSF and Minerva Foundation.

\setcitestyle{numbers}
\bibliography{turbulence1}	

\end{document}